\documentclass[aps,pra,twocolumn,superscriptaddress]{revtex4-1}

\usepackage{graphicx}
\usepackage{color,ulem}
\usepackage{amsmath}
\usepackage{amsfonts}
\usepackage{amssymb}
\usepackage{bm}
\usepackage{enumerate}
\usepackage{array}
\usepackage{etoolbox}
\usepackage{hyperref}

\hypersetup{colorlinks,linkcolor=blue,urlcolor=blue,citecolor=blue}

\newcommand{\ket}[1]{\ensuremath{|#1\rangle}}

\begin{document}

\title{Non-equilibrium dynamics in an optical transition from a neutral quantum dot to a correlated many-body state}

\author{F. Haupt} \email{fhaupt@phys.ethz.ch}
\homepage[]{\\http://www.quantumphotonics.ethz.ch}
\affiliation{Institute of Quantum Electronics, ETH Z\"{u}rich,
  CH-8093, Z\"{u}rich, Switzerland}

\author{S. Smolka}
\affiliation{Institute of Quantum Electronics, ETH Z\"{u}rich, CH-8093, Z\"{u}rich, Switzerland}

\author{M. Hanl}
\affiliation{Arnold Sommerfeld Center for Theoretical Physics, Ludwig-Maximilians-Universit\"{a}t M\"{u}nchen, D-80333 M\"{u}nchen, Germany}

\author{W. W\"uster}
\affiliation{Institute of Quantum Electronics, ETH Z\"{u}rich, CH-8093, Z\"{u}rich, Switzerland}

\author{J. Miguel-Sanchez}
\affiliation{Institute of Quantum Electronics, ETH Z\"{u}rich, CH-8093, Z\"{u}rich, Switzerland}

\author{A. Weichselbaum}
\affiliation{Arnold Sommerfeld Center for Theoretical Physics, Ludwig-Maximilians-Universit\"{a}t M\"{u}nchen, D-80333 M\"{u}nchen, Germany}

\author{J. von Delft}
\affiliation{Arnold Sommerfeld Center for Theoretical Physics, Ludwig-Maximilians-Universit\"{a}t M\"{u}nchen, D-80333 M\"{u}nchen, Germany}

\author{A. Imamoglu}
\affiliation{Institute of Quantum Electronics, ETH Z\"{u}rich, CH-8093, Z\"{u}rich, Switzerland}

\date{\today}

\begin{abstract}
We investigate the effect of many-body interactions on the optical absorption spectrum of a charge-tunable quantum dot coupled to a degenerate electron gas. A constructive Fano interference between an indirect path, associated with an intra dot exciton generation followed by tunneling, and a direct path, associated with the ionization of a valence-band quantum dot electron, ensures the visibility of the ensuing Fermi-edge singularity despite weak absorption strength. We find good agreement between experiment and renormalization group theory, but only when we generalize the Anderson impurity model to include a static hole and a dynamic dot-electron scattering potential. The latter highlights the fact that an optically active dot acts as a tunable quantum impurity, enabling the investigation of a new dynamic regime of Fermi-edge physics.
\end{abstract}


\maketitle

When a fermionic reservoir (FR) experiences a dynamically changing local perturbation, all its eigenstates are modified in response; the resulting Anderson's orthogonality catastrophe \cite{anderson1967infrared} plays a central role in the physics of quantum impurity systems. Along with the Kondo effect, \cite{GEIM1994,Frahm2006,Vdovin2007,latta2011quantum,tureci2011many} the most extensively studied quantum impurity problem is the Fermi edge singularity (FES): \cite{mahan1967excitons,NOZIERES1969,oliveira1985fano,OHTAKA1990} an optical absorption event induces a local quantum quench, causing dynamical changes in reservoir states that lead to power-law tails in the absorption line shape. This has been observed, for example, in the context of x-ray absorption in metals, \cite{NEDDERMEYER1976,ISHII1977,CALLCOTT1978a} where a large ensemble of deep-level states were ionized upon absorption and the resulting collective modification of the absorption line shape was measured. A related many-body effect has also been investigated in semiconductor structures incorporating a degenerate electron gas: \cite{MAHAN1967a,HAWRYLAK1991,KANE1994} In these studies, the modification of the absorption line shape is due to a rearrangement of the conduction-band electrons after the creation of an electrostatic potential by photoexcited quasimobile valence-band holes.

In this Rapid Communication, we report the observation of a FES due to a single localized hole in a charge-tunable quantum dot (QD) and a tunnel-coupled FR. In our experiments, the ionization of a QD valence-band electron takes place via two competing paths: (1) excitation of a QD neutral exciton followed by ionization due to tunneling of the conduction-band electron into the FR, and (2) a direct transition from QD valence band to an electronic state above the Fermi level of the FR. While in the classic x-ray absorption experiments only the latter process is relevant, in our experiments both contribute to single-photon absorption. Since both paths lead to final states of identical structure, involving a single-hole charged QD and a FR whose eigenstates are modified by the QD scattering potential, we observe a Fano interference \cite{fano1961effects}. Thanks to the constructive nature of this interference we can observe the signature of the FES despite the small transition probability associated with path (2). The presence of path (1) is also responsible for the dynamical local screening of the hole potential. Tuning the energy of the QD electron level with respect to the Fermi energy allows us to change the residual electron charge on the QD continuously, thereby varying the strength of the effective hole scattering potential. While in earlier optical experiments \cite{SKOLNICK1987,CHEN1990,CALLEJA1991,FRITZE1994,Yusa2000,Huard2000} a FES was observed by creating an undefined number of positive hole charges in the FR, we generate the electrostatic scattering potential by a single localized hole on a QD, defining a spatially well-isolated impurity. \cite{OHTAKA1990,Hawrylak1991a,kleemans2010many,Heyl2012} From resonant absorption measurements we can determine the dynamics of the potential scattering as well as the Fano parameters of the correlated many-body state.
\begin{figure*}[t] 
\centering
\includegraphics[width=1\textwidth]{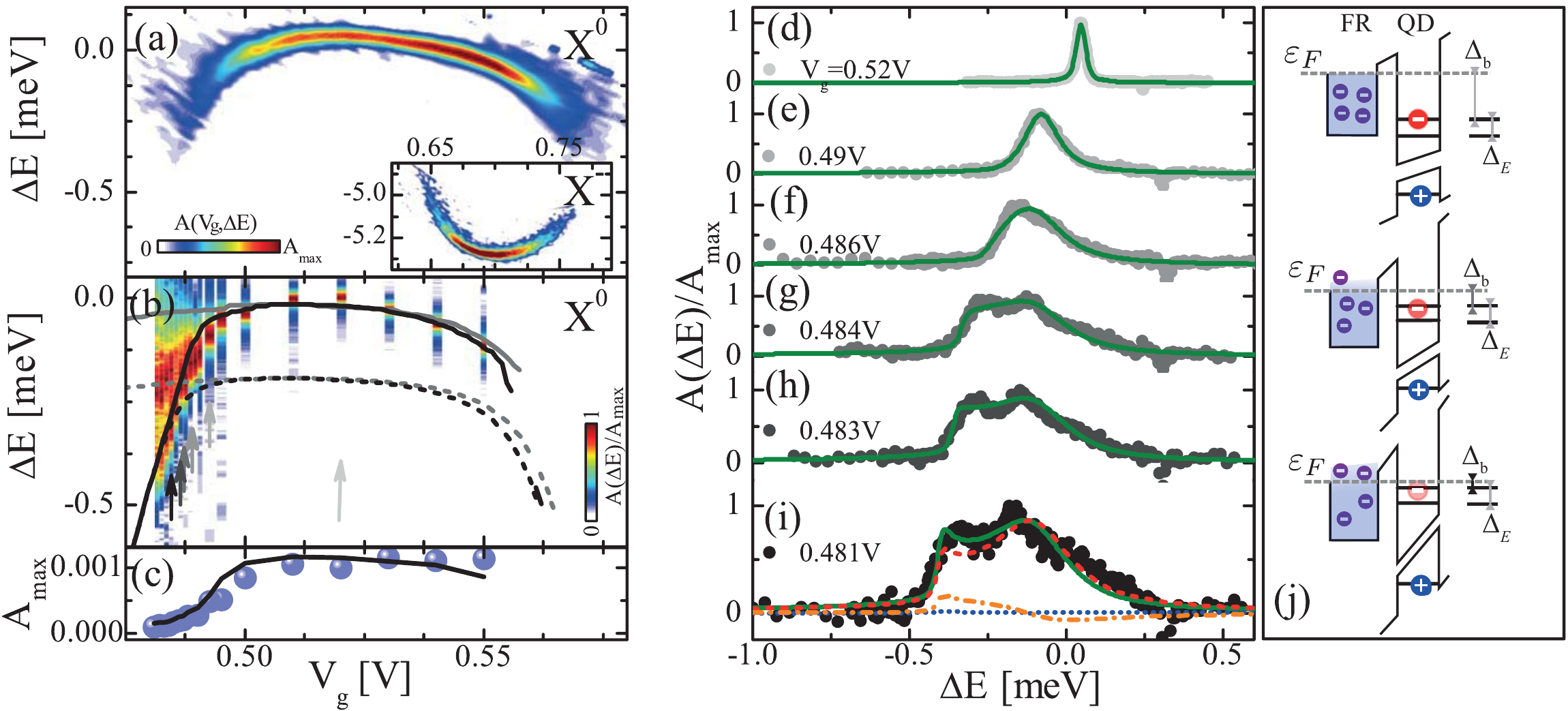}
\caption{(Color online) (a) Differential transmission measurements of the $X^0$ and the $X^-$ (inset) QD charging states, after subtracting the dc Stark shift (\cite{supple}). $\Delta E=E-E_0$ incorporates the peak absorption energy $E_0=1.3913\,$eV at $V_\text{g}=0.52\,$V. (b) High-resolution laser absorption scans (color scale) at selected gate voltages. Solid lines show fits of the calculated lowest-energy peak position, $\Delta E_{\rm peak}$, either without the scattering potential ($H_S^a = 0$, solid gray), or including it ($H_S^a \neq 0$, solid black). Dashed lines show the ground state energy difference, $\hbar \omega_{\rm th}$, between the initial and final states of the absorption process, calculated for $H_S^a = 0$ (dashed grey) or $H_S^a \neq 0$ (dashed black). The difference $\Delta E_{\rm peak} - \hbar \omega_{\rm th}$ is on the order of the dark-bright splitting $\Delta_E$ in the plateau center. (c) Comparison of the measured (dots) and calculated (curve) maximum absorption amplitudes (the latter scaled vertically by an overall fixed oscillator strength), shown as a function of gate voltage. (d)-(i) Measured absorption line shapes of the transition from a neutral exciton to a correlated many-body state (normalized by the experimental peak height $A_{\rm max}$) at gate voltages indicated by corresponding color-coded arrows in (b). The green curves display calculated results, scaled vertically and shifted horizontally to minimize the $\chi^2$ value of each fit (\cite{supple}). The absorption components of the direct (red dashed), indirect (blue dotted), and interference (orange dash-dotted) terms are exemplarily depicted in (i). Since the tail of the $X^0$ state spectrally overlaps with the $X^+$ state, we can excite the latter, which shows up as a dip in the absorption line shapes for red detunings. (j) Schematic of the renormalized transition energies of the bright ($\varepsilon_{F}-\Delta_{\rm b}$) and dark ($\varepsilon_{F}-\Delta_{\rm b}-\Delta_E$) electron levels with respect to the Fermi energy $\varepsilon_{\rm F}$ directly after the single-photon absorption event. $\Delta_{\rm b}$ indicates the energy difference between the Fermi energy and the bright state, corresponding to the line shape shown in (d) (top), in (f) (middle) and in (i) (bottom).}
\label{fig01}
\end{figure*}

\textit{Setup.---} The quantum impurity system under study, \cite{supple} consists of a single shallow self-assembled InAs QD with the neutral exciton resonance at $\lambda\approx891.25\,$nm, tunnel coupled to a $40\,$nm $n^\text{++}$ back gate and an In$_{0.08}$Ga$_{0.92}$As $7\,$nm quantum well that is $9\,$nm below the QDs. The system is embedded in a Schottky diode structure, in order to allow continuous tuning between different charging regimes.\cite{warburton2000optical,hogele2004voltage} Resonant laser spectroscopy measurements are carried out with a fiber-based confocal microscope setup (numerical aperture NA$=0.55$) that is embedded in a dilution refrigerator. Figure~\ref{fig01}(a) shows low-temperature differential transmission measurements of the energy plateaus of the neutral QD exciton ($X^0$) and single-negatively charged QD exciton ($X^-$) as a function of applied gate voltage. At the edges of the charging plateaus we observe an energy renormalization towards lower (higher) energies for the neutral (charged) QD transition, which is a hallmark of a strong tunnel coupling to a nearby FR. \cite{Dalgarno2008,latta2011quantum}

\textit{Measured absorption spectra.---} To probe the role of many-body interactions, we carried out high-resolution laser scans for various representative gate voltages in the $X^0$ plateau [Fig.~\ref{fig01}(b)]. Tuning the gate voltage to lower values allows us to increase the energy of the QD electron levels with respect to the Fermi energy. The absorption line shapes [$A(\Delta E)$] obtained for various gate voltages thus show the gradual evolution of the system from a regime where the final state is an excited QD state [Fig.~\ref{fig01}(d)] to the one in which it can be described by an optically excited electron in the FR and a hole trapped in the QD [Fig.~\ref{fig01}(i)]. We emphasize that for our sample the latter state has a dipole moment that is approximately a factor of 2 larger than the dipole moment of the $X^0$.

When the QD $X^0$ state approaches the Fermi energy, the absorption line shape consists of two peaks: the higher-energy peak corresponding to the $X^0$ transition that is tunnel broadened, and a second, lower-energy peak associated with the onset of absorption from the QD valence band directly into the FR [Fig.~\ref{fig01}(i)]. As we argue below, this second peak carries the signatures of a many-body resonance and reveals its nonequilibrium dynamics that is the focus of our work.

\begin{figure}[t]
  \centering
  \includegraphics[width=0.5\textwidth]{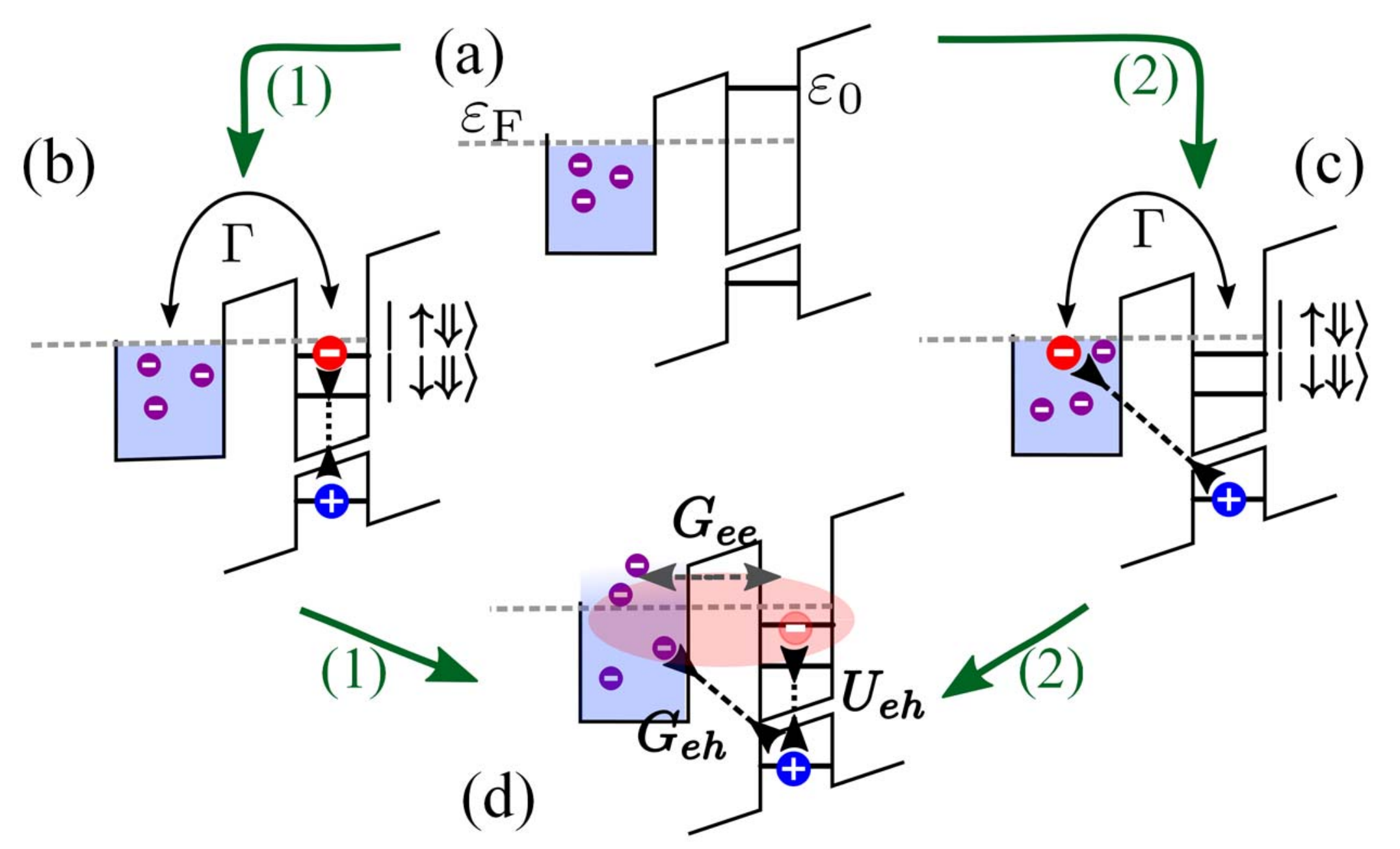}
  \caption{(Color online) Scheme of the quantum impurity system consisting of a single QD and a nearby FR. (a) Initial quantum state where the QD with energy $\varepsilon_0$ above the Fermi energy $\varepsilon_\text{F}$ is empty and the FR is unperturbed. The absorption of a single photon leads either (b) to a bound exciton on the QD or (c) to an indirect exciton. (d) The final state as $t \to \infty$ involves many-body correlations (red) between the FR and the QD. The black dashed and dotted lines depict the scattering potential between QD and FR, or the intradot Coulomb attraction, respectively.}
  \label{fig02}
\end{figure}

\textit{Model.---} In order to understand the various features of the absorption line shapes depicted in Figs.~\ref{fig01}(d)-\ref{fig01}(i), we generalize the excitonic Anderson model (EAM), previously used to describe the optical signatures of the Kondo effect,\cite{latta2011quantum,tureci2011many} by including a dynamic scattering potential:
\begin{eqnarray}\label{eq_total_H}
H_{A}^{a} &=& H^{a}_\text{QD}+H_\text{FR} +H^{a}_\text{S},\\
H^{a}_\text{QD} &=& \sum_{\sigma}\varepsilon ^a_\sigma(V_\text{g})\, \hat n_\sigma + U_\text{ee}\,\hat n_\uparrow\hat n_\downarrow +\delta_{a,f}\, \varepsilon_h(V_\text{g}),\\
H_\text{FR} &=& \sum_{k\sigma} \left[\varepsilon_{k\sigma}\,\hat c_{k\sigma}^{\dagger}\hat c_{k\sigma} + \sqrt{\Gamma/(\pi\,\rho)}\,(\hat e_\sigma^{\dagger}\hat c_{k\sigma} + \text{H.c.}) \right] \! ,\quad \phantom{.}\\
\label{eq_total_HS}
H^{a}_\text{S} &=&\Bigl[G_\text{ee}\,({\textstyle \sum_\sigma} \hat n_\sigma)- G_\text{eh}\,\delta_{a,f} \Bigr] {\textstyle \sum_{\sigma'}} \Bigl( \hat\Psi_{\sigma'}^\dagger\hat \Psi_{\sigma'} -{\textstyle \frac{1}{2}} \Bigr).\,
\end{eqnarray}
Here, $a=i,f$ differentiates between the initial~(i) Hamiltonian before absorption [Fig.~\ref{fig02}(a)], and the final~(f) Hamiltonian [Fig.~\ref{fig02}(d)] after creation of an exciton. In the QD Hamiltonian $H^{a}_\text{QD}$ the electron occupancy is denoted as $\hat n_{\sigma=\uparrow,\downarrow} = \hat e_{\sigma}^\dag \hat e_{\sigma}$. We assume a static hole spin and denote the bright state by $| \!\! \uparrow \Downarrow \rangle$ and the dark state by $|\!\! \downarrow \Downarrow \rangle$. \footnote{This assumption neglects the anisotropic exchange splitting between the bright states. In our description we fix the hole spin to be $\ket{\Downarrow}$ without loss of generality (\cite{supple})} The bare energy of the electronic level, measured with respect to the Fermi energy ($\varepsilon_\text{F}$=0), is given by $\varepsilon_\sigma^{a}(V_g) = \varepsilon_0(V_g)-\delta_{a,f} (U_\text{eh}+\delta_{\sigma,\downarrow} \Delta_E)$, where $\delta_{a,f}$ is the Kronecker delta and $\Delta_\text{E}$ is the dark-bright splitting. Both the conduction-band electron [$\varepsilon_0(V_g)=\varepsilon_0 - |e|V_g l$] and valence-band hole [$\varepsilon_h(V_g)=-\varepsilon_{h,0} + |e|V_g l$] state energies shift linearly with the gate voltage, $l$ being the voltage-to-energy conversion factor (lever arm). The energy of the optically excited QD states is lowered by the Coulomb attraction $U_\text{eh}$ and lifted by the on-site Coulomb repulsion $U_\text{ee}$. $H_\text{FR}$ describes the FR as a noninteracting conduction band with bandwidth $W$, symmetric around $\varepsilon_{\rm F}$, and constant density of states $\rho = 1/W$ per spin, tunnel coupled to the QD, where $\hat \Psi_{\sigma}=\sum_k \hat c_{k\sigma}$ annihilates a FR electron at the QD position and $\Gamma$ is the tunneling rate. Finally, the dynamic scattering potential $H^{a}_\text{S}$, which becomes important in the crossover between the local moment and free orbital regimes ($\varepsilon_\sigma^{f} \gtrsim 0)$,\cite{Heyl2012} describes the contact Coulomb attraction, $G_\text{eh}$, and repulsion, $G_\text{ee}$, between FR electrons and the QD hole or QD electrons, respectively, as depicted in Fig.~\ref{fig02}(d). Note that the effective scattering strength depends on the QD occupation and thus on the screened QD hole charge.

\textit{Fano interference.---} Starting from a neutral QD [Fig.~\ref{fig02}(a)], a photon absorption event can either create a QD exciton, involving $\hat e_\uparrow^\dagger$ [Fig.~\ref{fig02}(b)] or an indirect exciton, involving $\hat \Psi_{\uparrow}^\dagger$ [Fig.~\ref{fig02}(c)]. Both of these intermediate states evolve into a common final state [Fig.~\ref{fig02}(d)], where the QD hole scattering potential modifies the eigenstates of the FR due to the partial ionization of the QD and scattering of the FR electrons by the unscreened charge. The resulting absorption spectrum is given by
\begin{align}\label{eq_Abs}
\nonumber A( \nu)=& \alpha^2 \,A_\text{QD}(\nu) + \left(1-\alpha\right)^2 \, A_\text{FR}(\nu)\\
+& 2\, \alpha(1-\alpha)\,\cos(\phi)\, A_\text{I}(\nu),
\end{align}
with $\alpha$ being the branching ratio between the two optical paths and a Fano phase $\phi=0$ or $\phi=\pi$ \cite{supple}. $\nu=\omega - \omega_\text{th}$ describes the detuning between the laser frequency $\omega$ and the ground state energy difference, $\omega_\text{th}=(E^\text{f}_\text{G}-E^\text{i}_\text{G})/\hbar$, of the initial and final Hamiltonian. Using Fermi's golden rule, the direct absorption is calculated as $A_\text{QD}(\nu)= 2 \text{Re} \int _{0} ^\infty dt e^{i\nu t} \langle \hat e_{\uparrow}(t) \hat e_{\uparrow}^\dagger \rangle$ and the indirect Mahan absorption as $A_\text{FR}(\nu)= 2 \text{Re} \int _{0} ^\infty dt e^{i\nu t} \langle \hat \Psi_{\uparrow}(t) \hat \Psi_{\uparrow}^\dagger \rangle$ \cite{MAHAN1967a}. Here, we used the notation $\langle \hat b_2 (t) \hat b_1^\dagger \rangle=\text{Tr}(e^{i \bar{H}^i t} \hat b_2 e^{-i \bar{H}^f t} \hat b_1^\dagger \varrho)$, where $\hat b$ stands for either $\hat e_{\uparrow}$ or $\hat \Psi_{\uparrow}$, $\bar{H}^a=H^a-E^\text{a}_\text{G}$ and $\varrho$ is the Boltzmann weight at a FR temperature $T$.\cite{tureci2011many} The absorption line shape features a Fano interference, described by the term $A_\text{I}(\nu)= 2 \text{Re} \int _{0} ^\infty dt e^{i\nu t} \langle \hat \Psi_{\uparrow}(t) \hat e_\uparrow^\dagger \rangle$. The correspondence between the experimental [$A(\Delta E)$] and theoretical [$A(\nu)$] spectra follow from $\Delta E = \hbar \nu + \hbar \omega_{\rm th} - \tilde E_0$; here, $\tilde E_0$ is a fit parameter.\cite{supple}

\textit{Parameters.---} The recorded absorption maxima in Fig.~\ref{fig01}(b) are fitted with the calculated absorption maxima (black curves) that we obtained from a numerical renormalization group simulation \cite{Weichselbaum2007,*Weichselbaum2012} using Eq.~\eqref{eq_Abs}. A simultaneous fit of the charging plateaus and the $X^0$ line shapes allows us to extract all experimental parameters \cite{supple}. The intradot electron repulsion $U_\text{ee}=6.8\,\text{meV}$ is determined by the $X^-$ plateau length. From the $X^0-X^-$ energy separation, we extract $U_\text{eh}-U_\text{ee}=6.6\,\text{meV}$, neglecting correlation effects. In the center of the $X^0$ plateau [Fig.~\ref{fig01}(d)], the line width is determined by the FR-assisted relaxation into the dark exciton state, which in turn is determined by the gate voltage, the tunneling rate $\Gamma=400\,\mu$eV \cite{latta2011quantum}, and the dark-bright splitting $\Delta_E =175\,\mu$eV. The FR is characterized by its bandwidth $W=2\,\text{meV}$ and its temperature $T=120\,$mK. The best agreement between theory and experiment is obtained for $G_\text{eh}=\,3$meV and $G_\text{ee}=0.7 G_\text{eh}$. For comparison, we also plot the best fit of the $X^0$ plateau, if Coulomb scattering is ignored, i.e., $H^a_S\equiv 0$ [Fig.~\ref{fig01}(b) gray line]. As a result of the scattering potentials the lengths of the charging plateaus of $X^-$ and $X^0$ show different extents in gate voltage [Fig.~\ref{fig01}(b)]: This is in stark contrast to earlier experiments, which could be explained by assuming exclusively capacitive charging.\cite{supple} The renormalized energy of the final bright level with respect to the Fermi energy can be parametrized as $\tilde \varepsilon_\uparrow^f (V_g) = \varepsilon_0 (V_G) - U_\text{eh}+ \delta \varepsilon_0 (V_g)$, where $\delta \varepsilon_0 (V_g)$ accounts for a tunneling- and scattering-induced shift of the final bright level. Fitting model predictions to experimental data yields a lever arm of $l = 0.058$, $\varepsilon_0(0.52\,\text{V}) = 9.205\,$meV and $\tilde \varepsilon_\uparrow^f (0.52\,\text{V}) = -4.675\,$meV at $V_\text{g}=0.52\,$V.

\textit{Line shapes.---} The green curves in Figs.~\ref{fig01}(d)-\ref{fig01}(i) represent calculated absorption line shapes for the Hamiltonian [Eq.~\eqref{eq_total_H}] including the optical interference effect induced by the sample structure.\cite{supple} We highlight that we can only reproduce the experimental data using a Fano phase of $\phi=\pi$, corresponding to a constructive Fano interference between the direct and indirect transitions. $\alpha$ is determined by the square root of the ratio of the oscillator strengths of the direct and indirect transitions and is assumed to be independent of the exciton transition energy. In the present experiment we obtain the best agreement between experiment and theory for $1-\alpha=0.15$ \cite{supple}. Figure~\ref{fig01}(c) compares the measured maximum absorption amplitudes (dots) versus the calculated absorption amplitudes without adjusting any parameters. The agreement, up to a sample specific proportionality constant and fluctuations of peak contrast of the order of $10 \%$ due to alignment, underlines that our model reliably predicts the gate-voltage dependence of the peak absorption. The individual absorption line shapes of the direct (dashed curve), indirect (dotted curve) and interference (dash-dotted curve) terms are exemplarily shown in Fig.~\ref{fig01}(i). If the final neutral exciton levels are well below the Fermi energy [Fig.~\ref{fig01}(d)], the final state of the optical transition is the dark exciton state, which leads to a homogeneous broadening of the absorption line shape. In the tunneling regime, however, the final state is a correlated many-body state, which is a superposition of the FR states and the QD bright and dark exciton states. Close to the Fermi energy [Fig.~\ref{fig01}(i)], the final state has vanishing probability amplitude for finding the electron in the QD. In this regime the QD electron tunnels out into the FR lowering QD hole screening and thereby increasing the effective scattering potential. As a consequence, a screening cloud is formed in the FR that leads to a FES singularity. We emphasize that the absorption strength of the indirect element featuring the FES is very small ($\alpha =0.85$) and can only be detected due to a significant enhancement by the Fano interference. Due to the spectral overlap of $A_\text{FR}(\nu)$ and $A_\text{QD}(\nu)$, we cannot determine experimentally the power-law tail of the FES. However, the good agreement between our experimental data and theory indirectly demonstrates the presence of a FES.
\begin{figure}[t]
  \centering
  \includegraphics[width=0.5\textwidth]{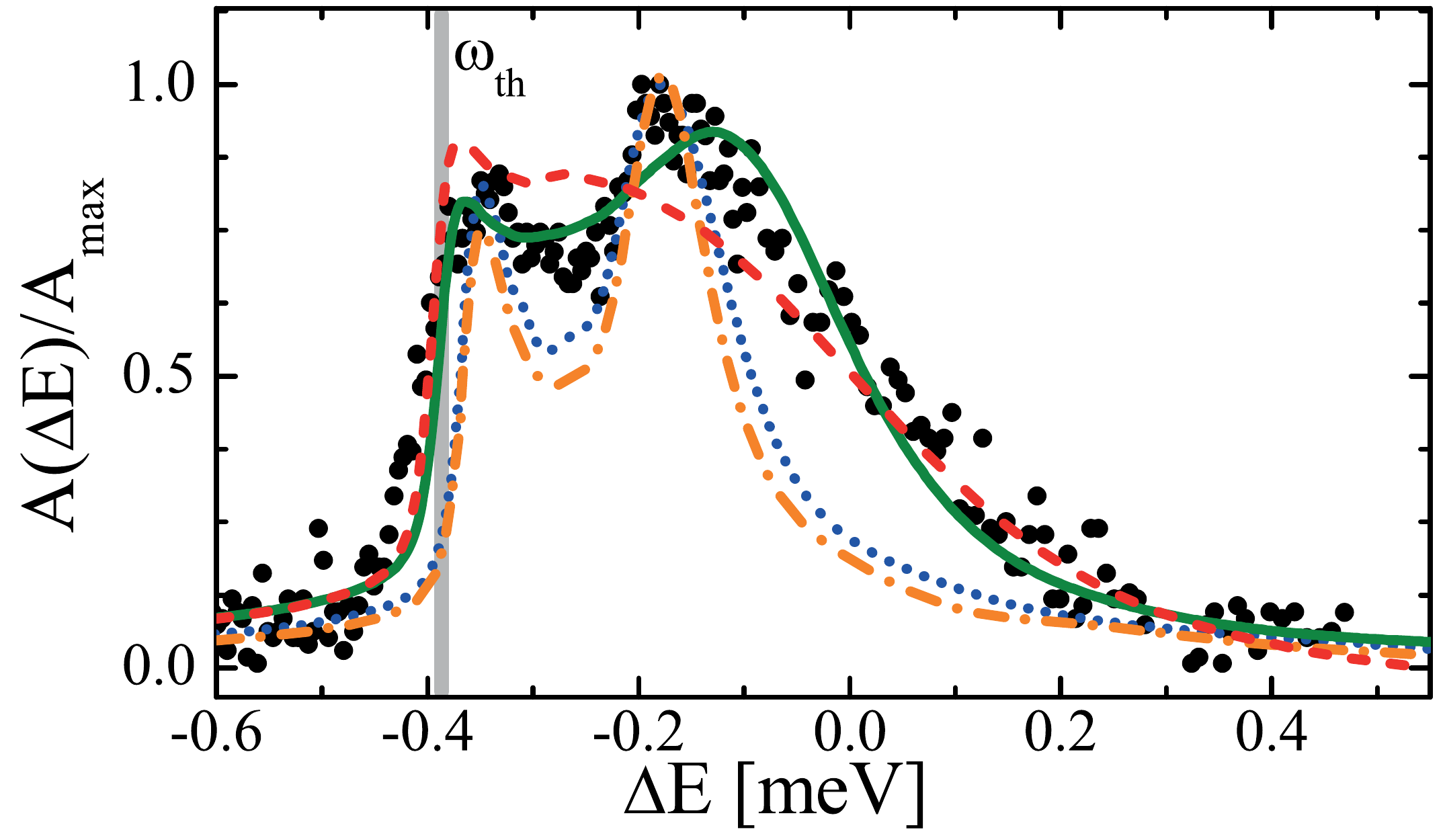}
  \caption{(Color online) Comparison of the experimental absorption line shape at $V_\text{g}=0.482\,$V with theory assuming different scattering scenarios. For the scales, the same conventions were used as for Figs.~\ref{fig01}(d)-\ref{fig01}(i). The red dashed line shows the best fit for the EAM model ($H^{a}_\text{S}=0$). Neglecting the electron-electron repulsion ($H^{a}_\text{S}\neq0$ with $G_\text{ee}=0$) the best fit yields the blue dotted curve, while the mean-field approach ($H^{a}_{\langle S \rangle}$) is shown by the orange dash-dotted curve. The green solid line depicts a dynamic scattering potential.}
  \label{fig03}
\end{figure}

\textit{Dynamical screening.---} In order to verify the role of the dynamical screening potential, we compare in Fig.~\ref{fig03} our experimental data with theory, for four different screening potentials. (i) The EAM model (dashed line, $H^{a}_\text{S}=0$) resembles the experimental data for $\nu \gg \Gamma$, indicating the absence of a scattering potential for very short time scales. As the indirect absorption spectrum $A_\text{FR}(\nu)$ only probes the constant density of states in the FR, the EAM model fails to reproduce the double-peak structure dominating the low-energy part of the spectrum. (ii) Inclusion of a scattering potential leads to the pronounced low-energy peak associated with the FES. Usually, the FES singularity is described by the Mahan-Nozieres-De Dominicis Hamiltonian,\cite{mahan1967excitons,oliveira1985fano,NOZIERES1969} which considers a scattering potential $G_\text{eh}$ while neglecting any Coulomb repulsion between QD and FR electrons, i.e. $G_\text{ee}=0$ (dotted curve). (iii) A possible way to include the latter interaction in our description while still using, for simplicity, a time-independent scattering potential would be to use a mean QD electron occupation, $H^{a}_{\langle S \rangle} = (G_\text{ee} \sum_{\sigma} \langle \hat n_\sigma(t \to \infty) \rangle -G_\text{eh}\delta_{x,f})\sum_{\sigma'} ( \hat\Psi_{\sigma'}^\dagger \hat \Psi_{\sigma'} -\frac{1}{2})$ (dash-dotted line). The models (ii) and (iii) both feature a second peak in the absorption spectrum in the long-time limit ($\nu \ll \Gamma$), but the absorption line shapes strongly deviate from the experimental data for short-time scales ($\nu \gg \Gamma$). (iv) Using the dynamical screening model of Eq.~(\ref{eq_total_HS}), we allow the QD electron occupation and thereby the screening of the QD hole potential to evolve in time. As depicted by the solid line, this dynamical screening model yields good agreement with experiment for all energy scales, showing that a scattering potential, and consequently an electron screening cloud in the FR, forms on time scales on the order of reciprocal $\Gamma$.

In contrast to prior nonresonant excitation experiments,\cite{kleemans2010many} we directly observe a correlated many-body state formed by the direct and indirect exciton transitions and develop a model to quantify the potential scattering strength. We note that our model assumes a perfect screening potential.\cite{HAWRYLAK1991} A partial screening of the scattering potential due to imperfections in the FR would lead to a stronger power-law decay of the FES,\cite{Gefen2002,Yusa2000,Abanin2004} which could explain the residual difference between experiment and theory. In conclusion, we demonstrated a \textit{dynamic} regime of Fermi-edge physics that highlights the importance of optically active quantum dots in the investigation of quantum impurity physics.

We acknowledge helpful discussions with B. Sbierski. This work was supported by an ERC Advanced Investigator Grant (F.H., S.S., W.W., J-M.S., A.I.). J.v.D., M.H., and A.W. acknowledge DFG via NIM, SFB631, SFB-TR12, and WE4819/1-1.

\pagebreak
\onecolumngrid
\appendix

The supplementary should provide background information to the studied effect. It is structured in two main sections. The first section deals with experimental methods and sample properties, e.g. the experimental setup and QD properties. The second section addresses the excitonic Anderson model with dynamic scattering potential in detail and states the parameterizations of the model.

\section{Experimental methods and sample properties}

\subsection{Experimental setup}
\begin{figure}[htp]\centering
 \includegraphics[width=0.65\textwidth]{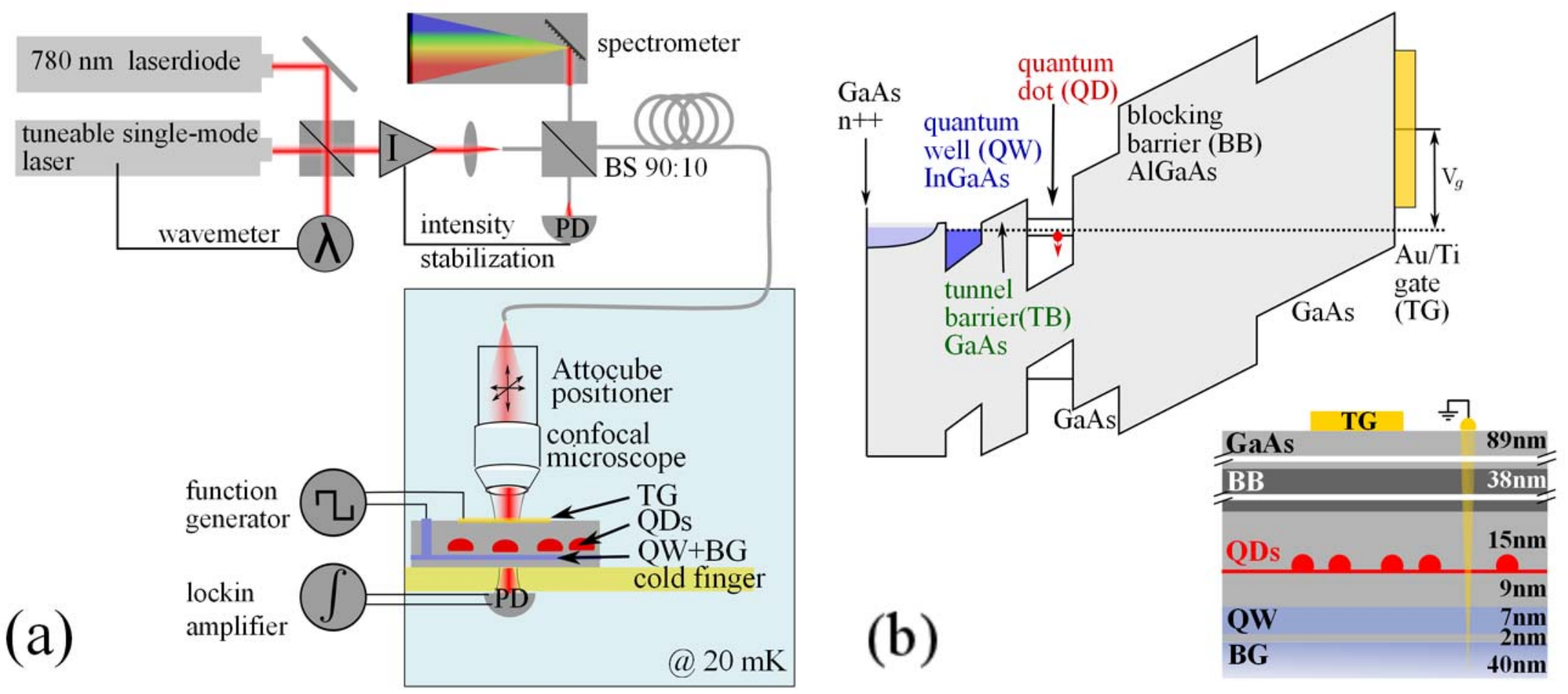}	
\caption{(a) Scheme of the experimental setup. The He$^3$/He$^4$ dilution refrigerator (blue shaded) has a fiber-based, confocal and movable objective. A $\lambda=780\,$nm laser diode, the 90:10 beam splitter (BS),  and the spectrometer are utilized for PL measurements. In differential transmission, we use a tunable laser. The sample is mounted on the cold finger of the cryostat and the resonantly scattered light of the QDs is detected by a photo diode (PD) mounted underneath the sample. (b) Schematic of the sample structure in the valence and conduction band picture (top) and growth structure (bottom). The sample consists of  a n++ back gate (BG), quantum well (QW), tunnel barrier(TB), QDs, blocking barrier (BB) and a Ti/Au top gate (TG). GaAs spacer layers are shown in grey. } \label{SetupSample} \end{figure}

The experimental setup, used to conduct the experiments presented in the main text, is schematically shown in Fig.~\ref{SetupSample}(a). In order to achieve cold temperatures, the sample is mounted inside a fiber-based confocal microscope embedded in a dilution refrigerator with a base temperature of $T=20\,$mK in the mixing chamber. X-Y-Z positioners on the microscope allow us to select individual QDs. The microscope objective has a numerical aperture of N.A.$=0.55$ featuring a diffraction-limited spot size. To conduct photoluminescence measurements (PL) we use a $\lambda=780\,$nm laser diode. The emitted PL signal is collected by the confocal microscope and spectrally analyzed with a spectrometer. As a second spectroscopy method, we perform differential transmission measurements by tuning a resonant single mode laser across the QD resonances, while the laser frequency and laser intensity ($P=0.87\,$nW) is stabilized against long-term drifts. To record the differential transmission signal, we modulate the resonance of the QD transition energies by varying the gate voltage between back gate and top gate with a frequency of $f=187\,$Hz and a modulation amplitude of $150\,$mV. For each laser frequency we detect the absorption difference signal with a Si photo diode that is mounted underneath the sample. The signal is afterwards analyzed with a lock-in amplifier.

\subsection{Sample structure}

\begin{figure}[htp]\centering
  \includegraphics[width=0.95\textwidth]{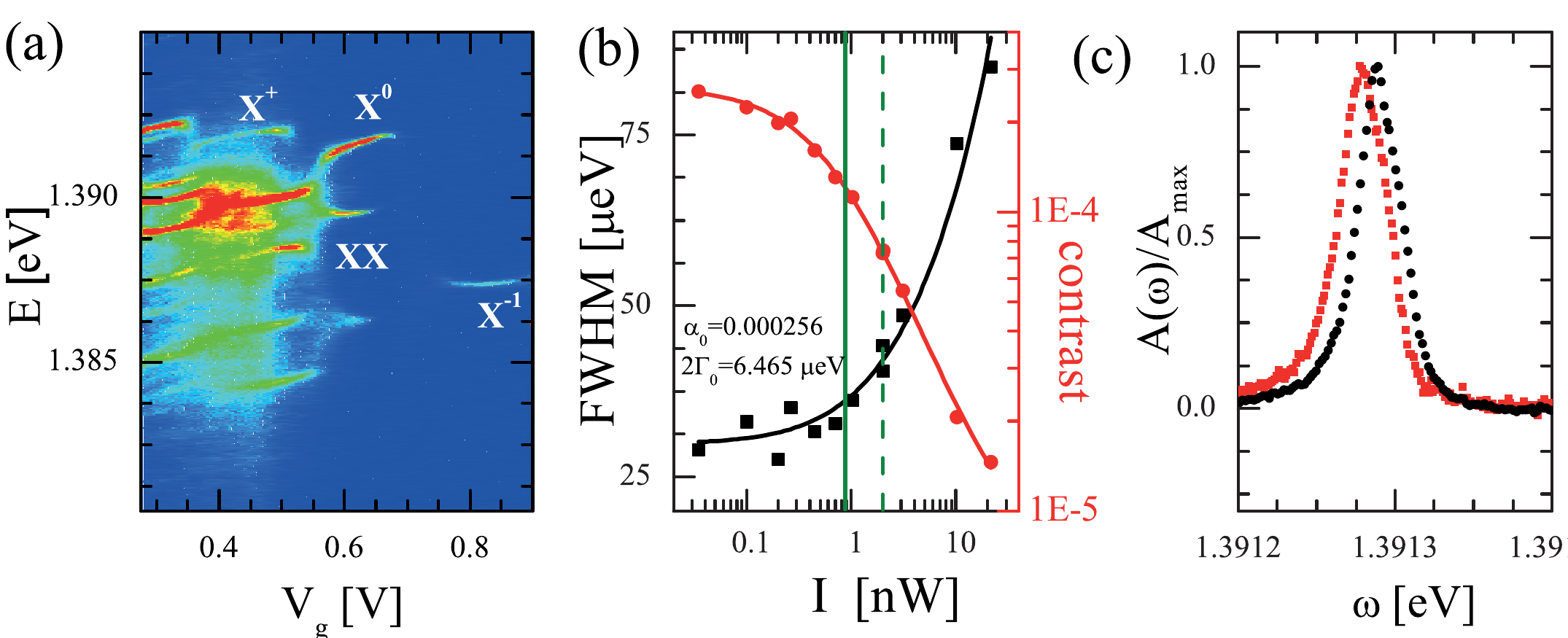}
  \caption{(a) Photoluminescence spectrum of the QD transition energies as a function of gate voltages, allowing us to identify the different charging plateaus. (b) Saturation behavior of the $X^0$ transition in the plateau center ($V_g=0.52\,$V). The full width at half maximum (FWHM, black squares) and the differential transmission contrast (red squares) are measured versus laser power. The solid lines are fits according to the experimental data \cite{Kroner2008_2}. The green solid line depicts the laser power used in the present experiments, while the dashed line shows the saturation laser power. (c) Spectral splitting of the $X^0$ transition at a gate voltage of $V_g=0.52\,$V. The $X^0$ is split at zero magnetic field in two bright states obeying different optical selection rules. The transition that we studied throughout the paper is depicted by filled black dots, whereas the energetically higher transition (red squares) is suppressed by adjusting the laser polarization.}
\label{XYsplittinSaturation}
\end{figure}

The sample structure consists of self-assembled InGaAs quantum dots (QDs) embedded in a Schottky diode structure, cf. Fig.~\ref{SetupSample}(b). The QDs are grown by molecular beam epitaxy. The sample design features a large tunnel coupling between the QDs and a nearby Fermi reservoir (FR). To this end, we used a modulation doped QW that defines a sharp boundary for the electron gas, consisting of a 40-nm-thick n++-doped GaAs layer and a In$_{0.08}$Ga$_{0.92}$As quantum well (QW), both coupled via a $2\,$nm GaAs barrier. The excitonic emission wavelength of the QW is $\lambda=840\,$nm. Due to segregation of dopants and the modulation doping, the QW/doped-layer system form the back contact (BC) of the Schottky diode. From transport measurements we obtain an electron density of $n=1.2\times 10^{12}\,\text{cm}^{-2}$. The tunnel barrier (TB) between QD and QW is designed to be $9\,$nm, which is in agreement with an estimate of the lever-arm, after fitting the experimental absorption line shapes, that suggests a TB of $8.5\,$nm. In order to prevent a current flow from the back gate to the semi-transparent 8 nm thick Ti-Au top gate, a blocking barrier (BB) of $38.5\,$nm Al$_{0.42}$Ga$_{0.58}$As is grown $15\,$nm above the QDs, cf. the valence band and conduction band diagram in Fig.~\ref{SetupSample}(b). Furthermore, the blocking barrier close to the QD serves to stabilize photo-excited holes in the QD \cite{ediger2007peculiar_2}.
Photoluminescence measurements of the QD emission as a function of gate voltage allows us to identify the different charging regimes [Fig.~\ref{XYsplittinSaturation}(a)]. In resonant QD spectroscopy, it is crucial to identify the saturation of an optical transition in order to prevent any power broadening \cite{Kroner2008_2}. In the presented experiments, we ensure that the laser power is 2.5 times smaller than the saturation power of~$\sim 2\,$nW, cf. Fig.~\ref{XYsplittinSaturation}(b).

\subsection{Fine structure splitting}

The neutral exciton $X^0$ is generated by a bound electron-hole pair having as eigenstates two bright states (denoted as $\ket{\uparrow \Downarrow} \pm \ket{\downarrow \Uparrow}$) and two dark states ($\ket{\downarrow \Downarrow} \pm \ket{\Uparrow \uparrow}$), that are split by exchange splitting ($H=-\sum_{i=x,y,z}(a_i \hat{J}_{h,i}\hat{S}_{e,i}+b_i \hat{J}_{h,i}^3\hat{S}_{e,i})$) \cite{Michler2003_2} at zero magnetic field. The bright and dark manifolds are split due to electron-hole exchange coupling ($\Delta_\text{E} \gg T$, see Sec.~\ref{param}), whereas the bright states are split by anisotropic exchange ($\delta_x$). After excitation of the bright exciton states, the QD electron can tunnel into the Fermi reservoir or decay into the dark state via co-tunneling events. As tunneling broadening ($\sim \Gamma$) is much larger than the anisotropic exchange splitting $\sim \delta \ll \Gamma$, the latter does not affect the absorption line shapes, when the bright states are close to or above $\varepsilon_\text{F}$ and is therefore negligible. In the center of the $X^0$ plateau, where tunneling is suppressed, we observe a polarization-dependent splitting of the bright excitons ($\delta_x=15\, \mu$eV), which is smaller than the line width, cf. Fig.~\ref{XYsplittinSaturation} (c) \cite{Gammon1996_2,Langbein2004_2}. This introduces a small uncertainty in the calculated NRG center frequency with respect to the experimental data for line shapes in the plateau center. As the hole spin-flip time is much longer than any other time scale of the system, we can fix the hole spin to be $\ket{\Downarrow}$ (or $\ket{\Uparrow}$ with equal probability). These assumptions allow us to treat the system by NRG using only the states $\ket{\uparrow \Downarrow}/\ket{\downarrow \Downarrow}$ (or $\ket{\downarrow \Uparrow}/\ket{\uparrow \Uparrow}$).

\subsection{Measurement of dc-Stark shift and dipole moment \label{DC}}

\begin{figure}[htp]\centering
  \includegraphics[width=0.9\textwidth]{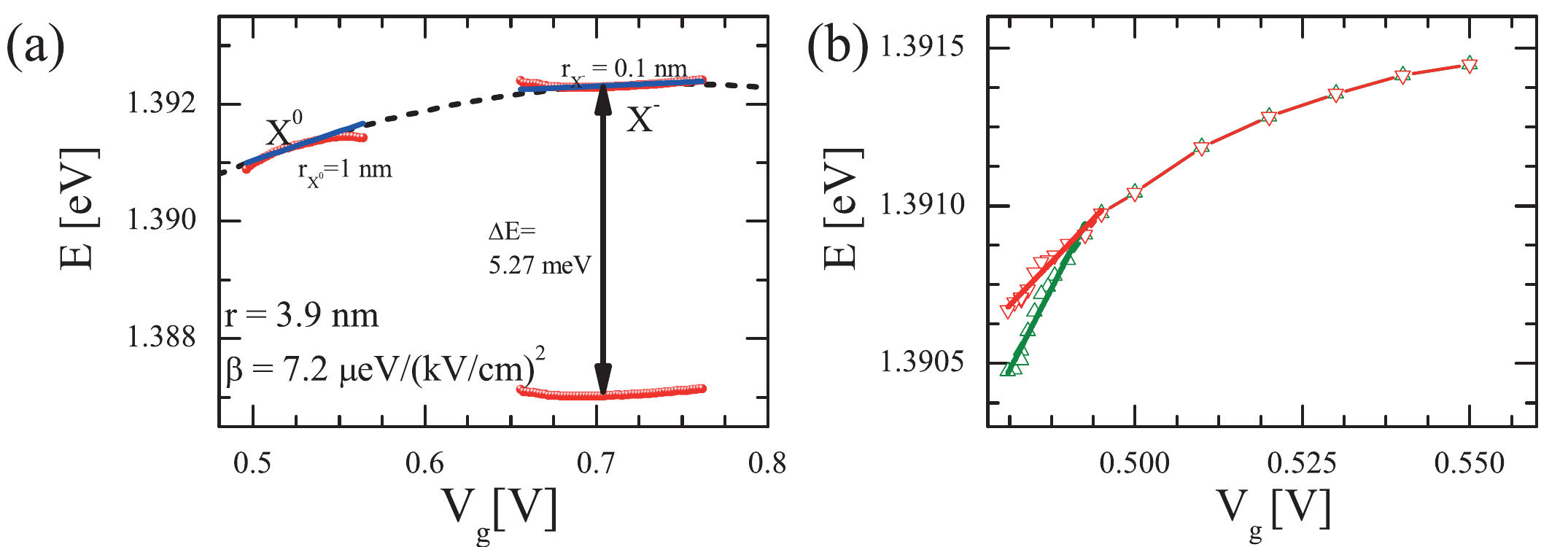}
\caption{(a) Quantum confined Stark shift of excitons in the QD. The experimental data (red bullets) show the maximum peak absorption positions of the $X^0$ and (shifted) $X^{-}$ charging plateaus that are fitted by a second order approximation of the Stark shift (dashed line). The linear approximations of the individual dc stark shifts for both charging states are shown as blue solid lines. The black arrow indicates the charging energy $\Delta E$ of the $X^-$ with respect to the $X^0$ transition. (b) Peak positions of the direct (red triangles) and indirect (green triangles) absorption peak maxima, plotted without subtracting the dc-Stark shift. The dipole moment of the indirect (green) transition is larger than that of the direct (red) transition by roughly a factor of 2.} \label{X0seperation} \end{figure}

The dc Stark shift in the present QD sample originating from the applied electric field can be approximated by two parameters, the permanent dipole moment $p$ and the polarizability $\beta$:
\begin{eqnarray}
\Delta E_\text{th}=E_\text{off}-p F+\beta F^2\, \text{with}\,  F=-\delta V_\text{g}/D,
\label{DCeq}
\end{eqnarray}
where $\delta V_\text{g}$ is the change of gate voltage with respect to the reference gate voltage $V_\text{g}=0.52\,$V. $D$ is the distance between the ohmic back contact and the metallic top-gate. The permanent dipole moment, $p$, is a measure of the spatial electron-hole separation in the exciton, $r=\frac{p}{e}$. In order to extract the relevant data we proceed as follows:
\begin{itemize}
  \item We approximate each charging plateau separately with a linear dc-Stark shift ($\tilde{E}_\text{th}=E_\text{off}-p F$) and obtain in the center of the $X^0$ ($X^-$) plateau an effective dipole moment $r_{X^0}$ at $V_g=0.52\,$V ($r_{X^-}$ at $V_g=0.71\,$V), cf. Fig.~\ref{X0seperation} (a).
  \item Then we extract the quadratic overall dc-Stark shift. Here, the previously determined individual linear dc-Stark shifts at $V_g=0.52\,$V ($V_g=0.71\,$V) serve as tangents with respect to the threshold value $E_\text{off}=1.3913\,$meV at $V_\text{g}=0.52\,$V. Using the charging energy $\Delta E$ of the $X^{-}$ with respect to the $X^0$ transition as another fit parameter in Eq.~\ref{DCeq}, we obtain the best fit with the highest confidence for $\Delta E=5.27\,$meV, cf. Fig~\ref{X0seperation}(a).
\end{itemize}
The experimental data of the $X^0$ and (shifted) $X^{-}$ charging plateaus show a permanent dipole of $r=3.9\, \text{nm}$ and a polarizability of $\beta =7.2\, \mu \text{eV}/ (\text{kV/cm})^2 $ \cite{Warburton2002_2}. These values are larger than for QDs with a low tunnel coupling, possibly because the wave function of the QD charges extends into the FR. The observed two-peak structure shows a gate-voltage dependent splitting, implying different dipole momenta for the direct and indirect excitons [Fig.~\ref{X0seperation}(b)].

Note, that the "bare" charging energy of the $X^-$ in the limit of low tunnel coupling ($\Gamma=0$) and no scattering potential $H_\text{S}=0$ is given by $U_\text{eh}-U_\text{ee}=6.6\,$meV provided that we ignore the correlation effects. As the QD levels hybridize with the FR due to tunneling events and the scattering potential, this leads to a energy renormalization  towards lower (higher) energies for the neutral (charged) QD transition ($\Delta E < U_\text{eh}-U_\text{ee}$). The energy renormalization difference is in the present experiment found to be $E_\text{r}=\Delta E -(U_\text{eh}-U_\text{ee})\sim -1.33\,$meV. For details on the determination of $U_\text{eh}$ and $E_\text{r}$ see Sec.~\ref{param}.

\subsection{Signatures of FES in other quantum dots}

\begin{figure}[htp]\centering
\includegraphics[width=0.95\textwidth]{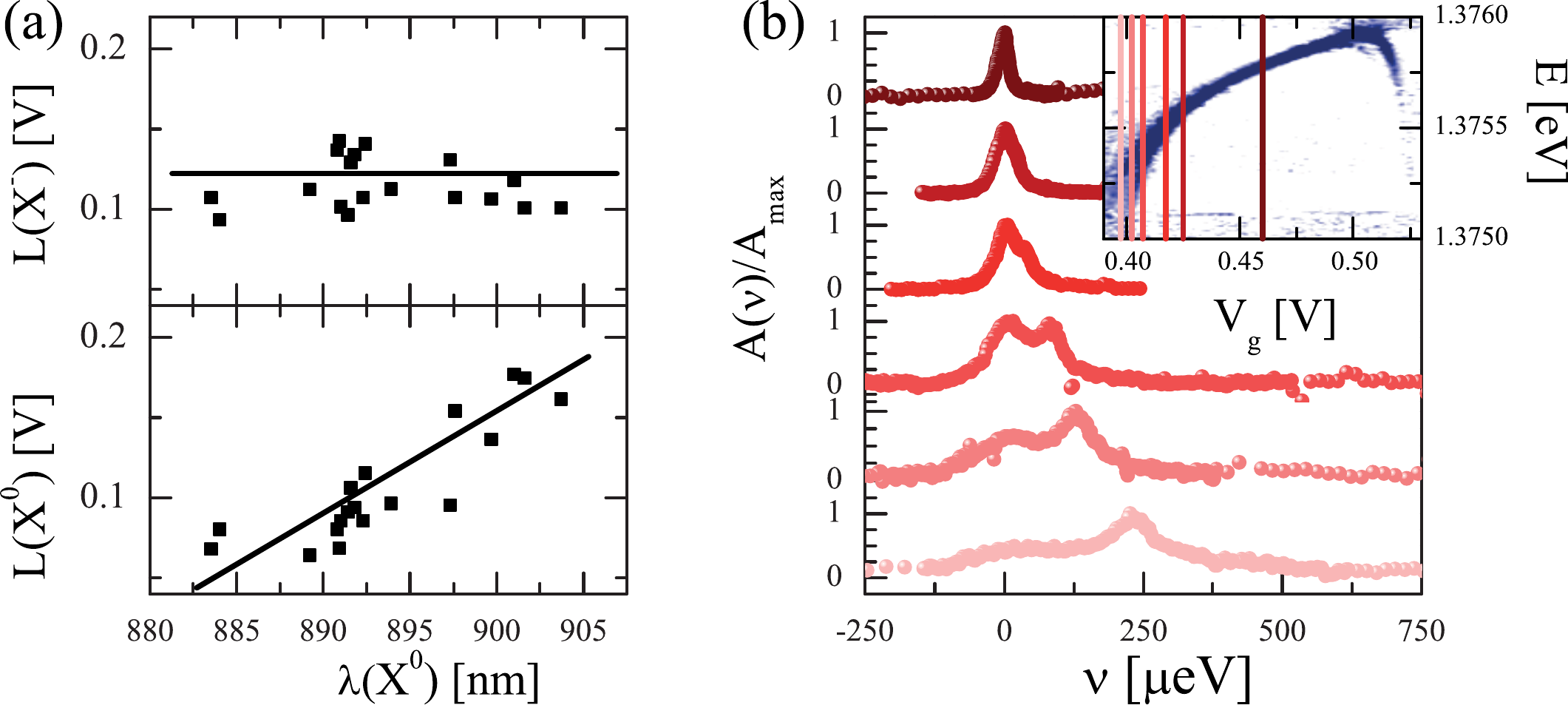}
\caption{(a) Gate voltage extents (lengths) of the $X^-$ and the $X^0$ plateaus of all studied QDs extracted from photoluminescence measurements. The black lines are a guide to the eye. (b) An example of the double peak structure in a different dot that has smaller tunnel coupling. Inset: 2D differential transmission scan of the $X^0$ plateau together with the indicated positions of the measured absorption line shapes.} \label{SampleStat} \end{figure}

We have studied the absorption line shapes of different self-assembled QDs. For QDs with a large tunnel-coupling we find a shortening of the $X^0$ charging plateau as well as a double peak structure. In  Fig.~\ref{SampleStat}(a), the plateau lengths of the $X^0$ and $X^-$ charging plateaus extracted from the PL measurements show a clear wavelength dependence. Generally, the charging regimes of QDs can be described using a capacitive charging model. If the trapping potential can be approximated by a parabolic potential, each charging plateau should have the same extent in gate voltage ($\propto U_{ee}$). However, we find that QDs at lower wavelengths have a significant shortening of the $X^0$ plateau length, $L(X^0)$, as compared to the $X^-$ plateau length, $L(X^-)$, suggesting that the dynamic scattering potentials (see main text) increase in strength for QDs with lower wavelengths.

As a further hallmark of the scattering potentials, a double peak like absorption line shape arises at the $X^0$ plateau edge which we observed for several different QDs. Figure~\ref{SampleStat}(b) shows a second QD with the $X^0$ at $E=1.3757\,$eV and a lower tunnel coupling $\Gamma$, which still shows the double peak structure.

\section{Parameterizations and Discussion of the Model}

\subsection{Parameterizations of the theoretical model \label{param}}

The extended Anderson model including a dynamic scattering potential, as discussed in the main text, can be parameterized with a single set of parameters by consecutive fitting of the $X^0$/$X^-$ plateaus and line shapes.

\begin{description}
  \item[$\Gamma=400\,\mu\text{eV}$, $\Delta_\text{E}=175\,\mu\text{eV}$] The charging plateaus experience a characteristic energy renormalization due to the tunnel-coupling $\Gamma$, as the QD electronic level hybridizes with the FR states and lowers the energy of the excited (ground) state of the $X^0$ ($X^-$). This results into a bending of the plateau edges towards lower (higher) energy. The line shape in the center of the $X^{0}$ plateau ($V_g=0.52\,$V) is determined by co-tunneling processes providing an irreversible decay into the dark state due to the dark-bright splitting ($T \ll \Delta_\text{E}$), which leads to homogeneous broadening. By fitting the data in the plateau center with theory, we estimate a dark-bright splitting of $\Delta_\text{E}=175\,\mu$eV, which is in agreement with prior experiments, that measured a dark-bright  splitting of $200-500\, \mu$eV \cite{Smith2005_2}. To minimize the number of variables, we have chosen $\Gamma=400\,\mu$eV from the center of the $X^0$. It is plausible that $\Gamma$ for the $X^{-}$ is larger. Better fits could be obtained for letting $\Gamma$ vary as a function of gate voltage $V_\text{g}$.
  \item[$U_\text{ee}=6.8\,\text{meV}$] The QD electron-electron repulsion $U_\text{ee}$ is determined by the plateau length of the $X^-$.
  \item[$G_\text{ee}=2.1\,\text{meV}$, $G_\text{eh}=3.0\,\text{meV}$]   As shown in the main text, the difference of the $X^0$ and $X^-$ plateau length can be explained by a dynamic scattering potential. As $G_\text{ee} \leq G_\text{eh}$, the $X^0$ charging plateau is fitted best using $G_\text{eh}=G_\text{ee}/0.7=3\,$meV. The accuracy in the fits of the absorption line shapes could be improved by varying $G_\text{eh}$ and $G_\text{ee}$ throughout the charging plateau, which was not done here.
  \item[$W=2.0\text{meV}$] The bandwidth of the Fermi reservoir was fitted from the line width of the absorption line shapes in tunnel-regime.
  \item [$U_\text{eh}=13.35\,\text{meV}$] The QD electron-hole attraction $U_\text{eh}$ is extracted as follows: the charging energy $\Delta E=5.27\,$meV between the centers of $X^-$ and $X^0$ plateaus has the form $\Delta E = U_\text{eh}-U_\text{ee}+E_\text{r}$, where $E_\text{r}$ is the shift accounting for the effects of level hybridization and the scattering potential $H_\text{S}$, cf. Fig.~\ref{X0seperation}(a). Using the above-mentioned value for $G_\text{eh}$, $G_\text{ee}$, $\Gamma$ and $W$, our NRG calculations yield a shift of $E_\text{r}=-1.33\,$meV (its value is dominated by scattering, since the tunneling contribution at the plateau centers is rather small). As a result, we deduce $U_\text{eh}=\Delta E + U_\text{ee}- E_\text{r}=13.35\,$meV. Note, we neglected any effect due to correlations, which could lead to correction of the exact value of $U_\text{eh}$ and $U_\text{ee}$.
  \item[$l=0.058$] The lever arm is $l=\tilde{D}/(D+\tilde{D})$, where $\tilde{D}=8.5\,$nm is the QD/FR distance and $D=138.5\,$nm is the QD/top gate distance indicated by both NRG calculation and the PL emission. This parameters can differ by a few \AA$\,$ from the predicted growth parameter ($\tilde{D}=9\,$nm,$D=138.5\,$nm) due to finite accuracy in the growth of the sample.
  \item[$T=120\,\text{mK}$ ] The temperature $T$ of the FR electrons is extracted by the red tail of $X^-$ line shapes (see Sec.~\ref{Temp}).
  \item[$\alpha=0.85$, $\phi=\pi$] The branching ratio and Fano phase can be fitted for the cases in the tunnel regime (see Sec.~\ref{brancha}).
\end{description}

\subsection{\label{Temp} Optical interference and sample temperature}

In the following we discuss an optical interference effect caused by the sample structure \cite{warburton2000optical_2} using absorption measurements of the single electron charged exciton $X^-$. The incident laser field on the sample $E_L$ is Rayleigh scattered at the QD at which the scattered light can be forward and backward scattered. The backward scattered (reflected) light travels in GaAs (refractive index $n=3.55$) a distance $D=138.5\,$nm to the top gate and will be again partly reflected at the top-gate (reflectivity $r=0.75$, fitted). In first approximation the field at the transmission detector [see Fig.~\ref{SetupSample}(b)] is a superposition of the transmitted light and all scattered components. Here, the back-scattered light accumulates a phase due to a different path length of $\varphi=\frac{2\pi n}{\lambda}2D$:
\begin{eqnarray}
E_{tot}=E_L e^{i \pi/2}+\chi(\lambda)E_L+r e^{i\varphi}\chi(\lambda)E_L=i E_L\left[ 1-i\chi(\lambda)(1+r e^{i\varphi}) \right],
\end{eqnarray}
with $\chi$ being the susceptibility of the QD with absorption $\Im(\chi)$ and dispersive part $\Re(\chi)$. Due to employing a Gaussian shape on our beam, the laser field $E_L$ acquires a Guoy phase $e^{i \pi/2}$ in the far field regime. Furthermore, in a differential transmission lock-in-method, the QD response function is modulated at a certain frequency such that the measured absorption is at this frequency:
\begin{eqnarray}
A(\lambda)=\Delta I/I\sim\Re\left[ i \chi(\lambda) (1+r e^{i\varphi}) \right],
\label{absorption}
\end{eqnarray}
where $I=E_{tot}^2$ is the intensity detected by the photo detector and we neglected terms $\sim (\chi(\lambda))^2$ as $1\gg \chi(\lambda)$. The back-scattered light mixes the dispersive part with the absorption part of the forward-scattered light, which leads to an optical interference. The numerical renormalization group (NRG) theory used in this paper for the calculation of the absorption line shapes $A_{NRG}$ considers $\Im(\chi_\text{NRG}(\lambda))=-A_\text{NRG}(\lambda)$. In order to calculate the dispersive response and thereby incorporate the optical interference we convolute the calculated absorption spectra using the Kramers-Kronig relation $\Re(\chi_\text{NRG}(\lambda))=-\frac{1}{\pi} {\cal P}\int{ d\lambda' \frac{\Im(\chi_\text{NRG}(\lambda'))}{\lambda-\lambda'}}$.
\begin{figure}[htp]\centering
\includegraphics[width=0.87\textwidth]{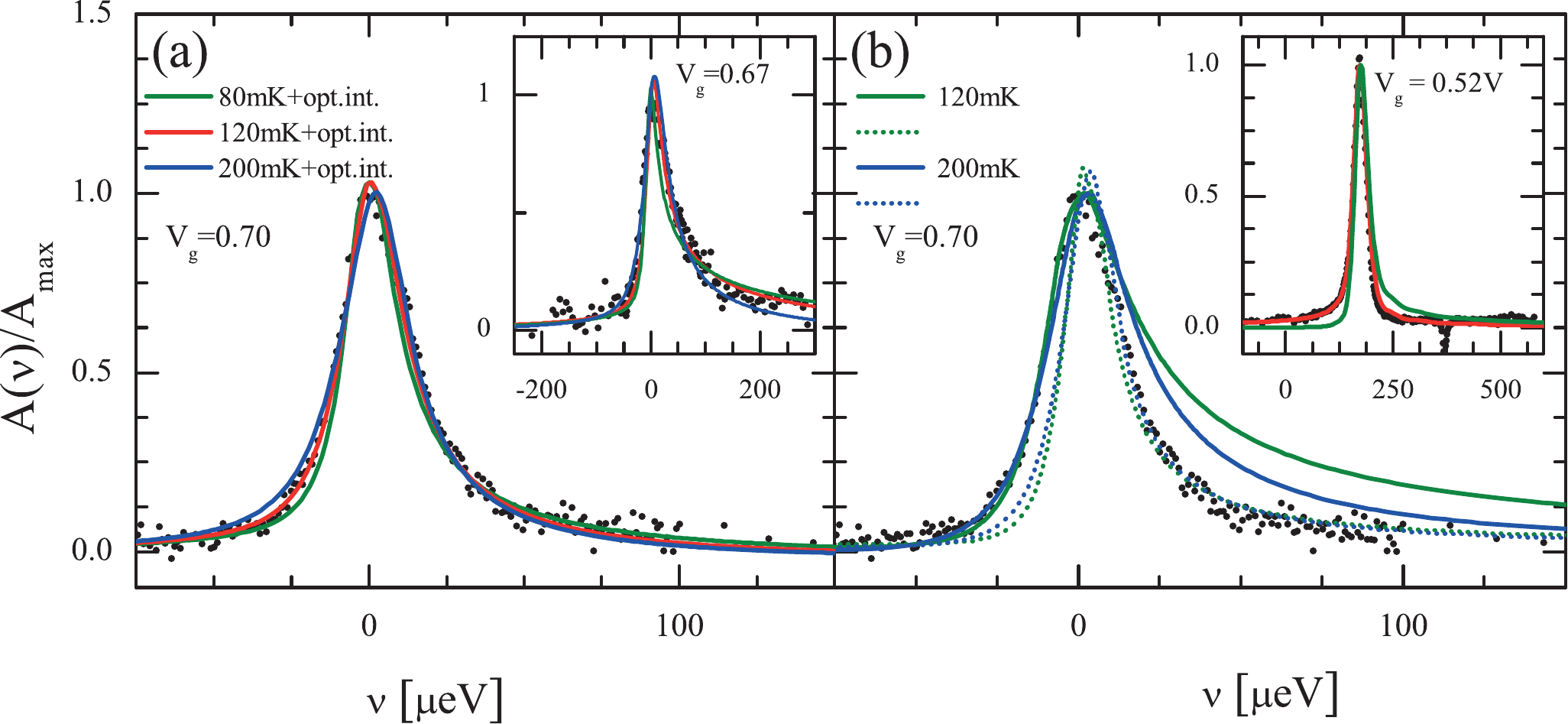}
\caption{(a) Fits of $X^-$ line shapes at $V_g=0.70\,$V (inset at $V_g=0.67\,$V) for different temperatures, i.e., $T=80\,$mK (green), $T=120\,$mK (red) and $T=200\,$mK (blue) including the optical interference (without scattering potential $H_S = 0$). (b) Fits of the $X^-$ absorption line shape ($V_g=0.70\,$V) without accounting for the optical interference. The dotted line focusses on fitting the blue tail of the experimental data, whereas the solid lines fits the red tail. The inset shows the role of optical interference for a $X^0$ absorption line shape at $V_g=0.52\,$V by incorporating  (red curve) and neglecting (green curve) the optical interference.} \label{Interference} \end{figure}

The electron occupation of the fermionic reservoir (FR) is governed by a Fermi-Dirac distribution. Shake-up processes due to finite temperature modify the available states in the Fermi reservoir. For the negative charged exciton $X^-$ of a highly tunnel-coupled QD, the absorption tails for red detunings ($\nu < -T)$ are governed by the levels below the Fermi energy in the FR. The line shape thus shows an exponential tail that depends on temperature $\sim e^{-\frac{\nu}{k_B T}}$ \cite{latta2011quantum_2}. For $\nu \ll \Gamma$, the line shape is governed by the spontaneous emission and resembles a Lorentzian tail. From fits of the red tails of $X^-$ line shapes with NRG calculations including the optical interference effect at $V_g=0.67\,$V and $V_g=0.70\,$V [Fig.~\ref{Interference}(a)], we extract a FR electron temperature of $T=120\,$mK. For comparison we show in Fig.~\ref{Interference}(a) fits for different temperatures. In Fig.~\ref{Interference}(b), we demonstrate that without the optical interference we cannot reproduce the full absorption line shape. Either we can fit the high energy or the low energy tail of the line shape. As the $X^0$ red tail is distorted by the dark-bright splitting, we determined the parameters for the optical interference and the FR temperature from the $X^-$ line shapes to calculate the $X^0$ line shapes, which are in good agreement with the experiment,  c.f. the inset of Fig.~\ref{Interference}(b) (green curve shows the bare NRG calculations; red curve includes the optical interference).

\subsection{\label{brancha} Branching ratio $\alpha$ between the direct and indirect transition}

\begin{figure}[htp]\centering
  \includegraphics[width=0.98\textwidth]{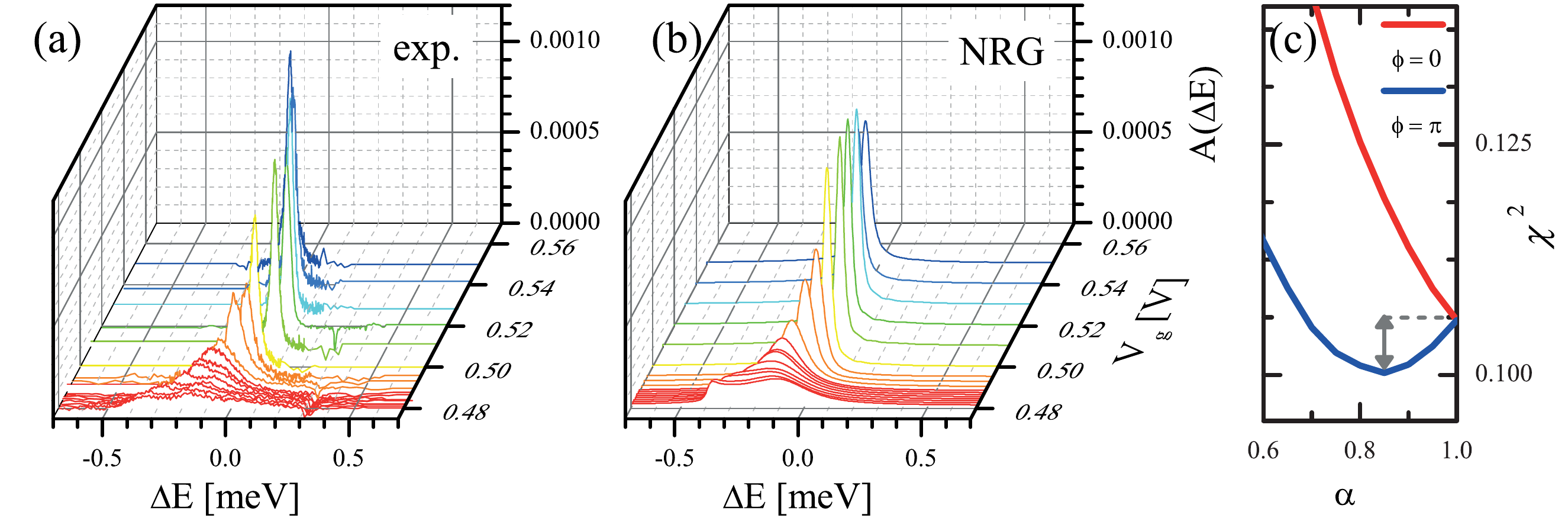}
\caption{Comparison of measured line shapes (a) and NRG simulations (b) in the gate voltage range of $V_\text{g}=0.48\,$V to $0.55\,$V (not normalized, cf. Fig. 1(c)) (c) Calculated cumulative $\chi^2$ of all measured line shapes from $V_\text{g}=0.481\,$V to $0.55\,$V for destructive interference ($\phi = 0$, red) and constructive interference ($\phi = \pi$, blue) as a function of branching ratio $\alpha$.} \label{chi2} \end{figure}

The branching ratio between the direct and the indirect transition can be determined by fitting the experimental data [Fig.~\ref{chi2} (a)] with NRG calculations [Fig.~\ref{chi2} (b)]. It is defined by the wave function overlap of the QD electron and QD hole versus the FR electrons and the QD hole. This ratio is specific for every QD. Since at zero magnetic field the Hamiltonian can be chosen to be real, the Fano phase has to be $\phi=0$ or $\phi=\pi$.

Due to variations in the peak absorption contrast stemming from the fact that the experiments were carried out on different days with uncontrolled changes in the alignment (on the order of $10\%$), we determined the values of the branching ratio $\alpha$ and the relative phase $\phi$ applying the following fitting procedure. This allows us to quantify and minimize the disagreement between theory and experiment as a function of $\alpha$ and $\phi$ for about $2700$ data points in total:

Using the system parameters of the Hamiltonian extracted as described in Sec.~\ref{param} we generate a set of initial and final eigenstates by diagonalizing the initial and final Hamiltonians using NRG. With Fermi's Golden rule we calculate the absorption spectra terms: $A_\text{QD}$, $A_\text{FR}$, and $A_\text{I}$. The theoretical line shape is then given by $A^\text{NRG}(\omega) = \alpha^2 A_\text{QD}(\omega)+ (1-\alpha)^2 A_\text{FR}(\omega) + 2 (1-\alpha)\alpha A_\text{I}(\omega) \cos(\phi)$, where $\omega = \nu + \omega_\text{th}$, with $\nu$ being the detuning between the laser frequency, $\omega$, and the absorption threshold energy, $\omega_\text{th} = (E^\text{f}_\text{G}-E^\text{i}_\text{G})/\hbar$, that is given by the ground state energy difference between the final and initial Hamiltonian. To compare the calculated and the measured line shapes for a given gate voltage, we fit the theoretical curve $aA^\text{NRG}(\nu + \omega_\text{th} - \tilde{E_0})$  to the normalized experimental data $A^\text{exp}(E-E_0)/A_\text{max}$, where $E_0 = 1.3913\,$eV is the peak absorption energy at $V_\text{g} = 0.52\,$V. $\tilde{E_0}$ and $a$ are curve-specific fit parameters, relating to the peak height and peak position. The fitting proceeds by first fixing $\{\alpha,\phi\}$ and then varying $\{a,\tilde{E_0}\}$ for each gate voltage to minimize the $\chi^2$. Changes in the electromagnetic environment upon optical excitation lead to random charging events of the nearby defects and/or charge accumulation at the AlGaAs/GaAs interface, which in turn modify the electric-field seen by the quantum dot at a given gate voltage. This results in variations of the experimental peak position of $\sigma(\tilde{E_0})=30\,\mu$eV with respect to the theoretical predictions, c.f. Fig.~1(b). $a$ is constant for each absorption line shape and accounts for the day-to-day variations in the sample alignment which in turn leads to a modified absorption contrast (i.e. the area under the absorption line shape). For completeness, we show in Fig. 1(c) of the main text the measured experimental peak contrast together with the pure peak absorption predicted by the NRG calculation (without compensating for alignment issues). From the deviations between experiment and theory we estimate the unavoidable variations in oscillator strength to be on the order of $10\%$. As an objective measure of the quality of the fit we calculate the $\chi^2$ value for each line shape as a function of the branching ratio $\alpha$ (for constructive, $\phi = \pi$, and destructive interference, $\phi = 0$, separately). Afterwards we estimate the cumulative $\chi^2$ of all measured (in total 18) line shapes as a function of $\alpha$. The best overall fit in the range of $V_\text{g}=0.481\,$V $- 0.55\,$V is obtained for $\alpha=0.85$ and $\phi=\pi$ [Fig.~\ref{chi2} (c)]. The case of destructive interference ($\phi=0$) shows an optimal branching ratio of $\alpha=1$ for all line shapes, but the corresponding $\chi^2$ is larger than the lowest $\chi^2$ obtained for $\phi=\pi$.

\begin{figure}[htp]\centering
  \includegraphics[width=1\textwidth]{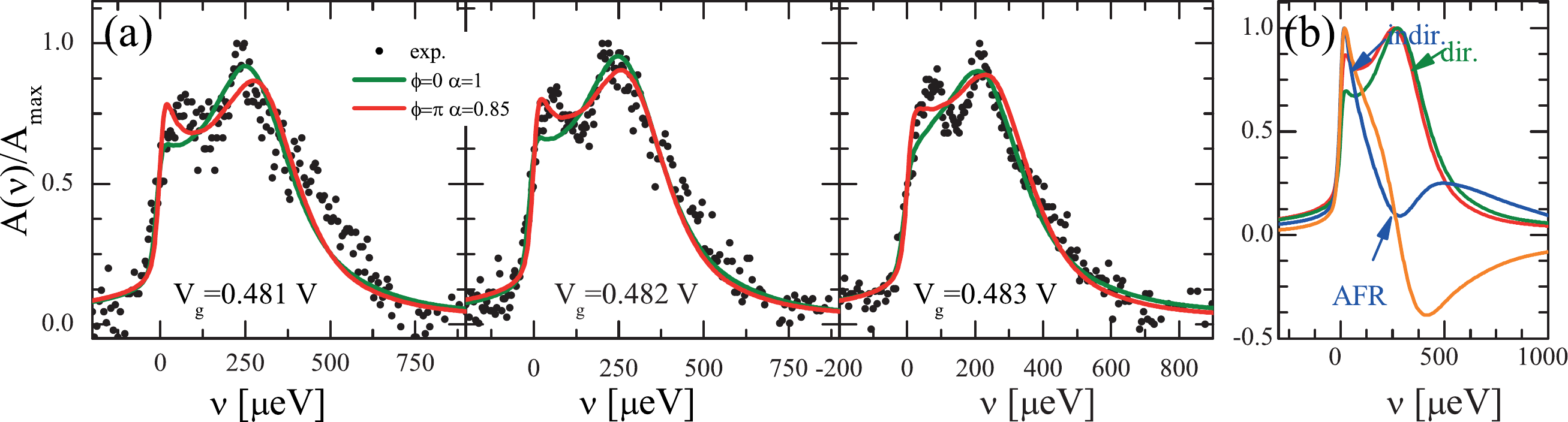}
\caption{(a) Comparison of line shapes for $V_g=0.481\,$V, $V_g=0.482\,$V and $V_g=0.483\,$V for the best fitting values of branching ratio $\alpha$ for constructive ($\phi=\pi$, red) and destructive ($\phi=0$, green) Fano interference. (b) Calculation of the decomposed normalized direct (green), indirect (blue) and interference (orange) absorption spectra for $V_g=0.482\,$V. In the indirect transition an Anti-Fano resonance (AFR) can be observed.} \label{Antifano} \end{figure}

In Fig.~\ref{Antifano} (a), we compare the theoretical predictions for constructive and destructive interference at $V_\text{g}=0.481\,$V, $0.482\,$V and $0.483\,$V. We emphasize that only the theoretical curve using a constructive interference can reproduce the low-energy peak associated with the Fermi edge singularity [Fig.~\ref{Antifano}]. Oliveira et al. \cite{oliveira1985fano_2} predicted an Anti-Fano resonance if the branching ratio $\alpha$ is equal to zero, which corresponds to an exclusively indirect exciton absorption. This scenario cannot be achieved in the present experiments since the QD states (direct transitions) always have a non-vanishing oscillator strength. However, in the NRG calculations, where we can decompose the individual contributions to the absorption line shape, we observe the Anti-Fano resonance (AFR) in the indirect transition [see blue curve in Fig.~\ref{Antifano}(b)].

\subsection{Comparison of the charging plateaus with theory and estimation of electron occupations \label{NRGsection}}

\begin{figure}[htp]\centering
  \includegraphics[width=1\textwidth]{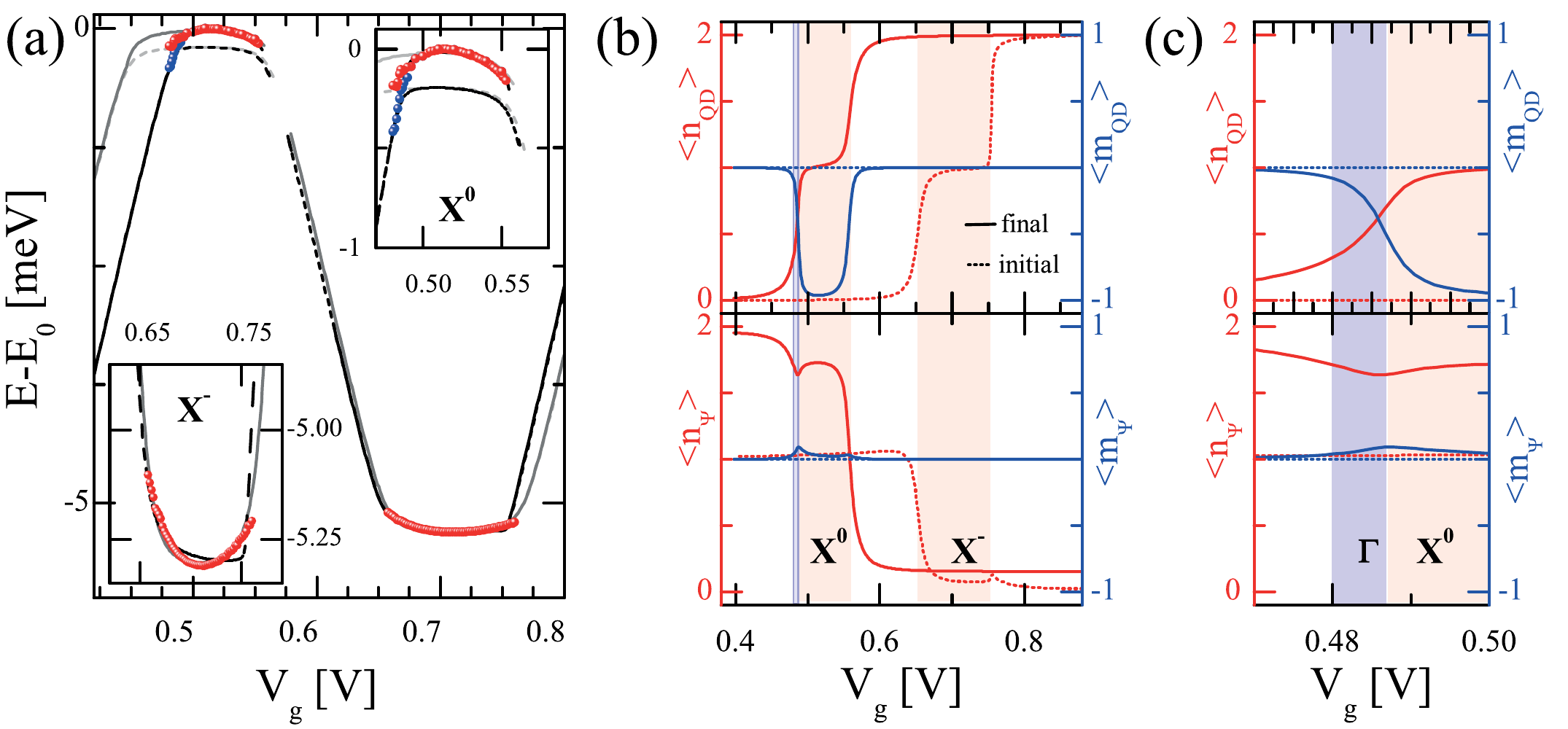}
\caption{(a) Fits of the measured charging plateaus using NRG calculations together with a zoom-in of the $X^0$ (inset upper right) and $X^{-}$ (inset lower left) using the extracted parameter stated in the paper for the $H^{a}_\text{S} =0$ (grey) and $H^{a}_\text{S} \neq 0$ (black). The solid lines depict the maximum absorption strength and the dashed lines are the ground state energy differences of the initial and final states of the system. (b) Occupation ($\langle n_x \rangle=\langle n_{\uparrow,x} \rangle+ \langle n_{\downarrow,x} \rangle$, red) and population inversion of the $\ket{\uparrow}$ and $\ket{\downarrow}$ state ($\langle m_x\rangle=\langle n_{\uparrow,x}\rangle-\langle n_{\downarrow,x}\rangle$, blue) of the QD ($x=\text{QD}$) and of the FR at the position of the QD ($x=\Psi$) for the initial (dotted lines) and the final state (solid lines) including the dynamic scattering potential. (c) Zoom-in of the tunnel-regime of (b).} \label{densityNORMAL} \end{figure}

In the following we discuss the fits of the experimentally measured charging plateaus as well as the electron occupations in the QD and the Fermi reservoir using a Numerical Renormalization Group (NRG) approach. The results are displayed in Fig.~\ref{densityNORMAL}. While at QDs without a scattering potential the $X^0$ and $X^{-}$ charging plateaus have equal lengths, we find a different result (experimentally and theoretically) as the scattering potential $G_{eh} \neq G_{ee}$ is turned on [Fig.~\ref{SampleStat}(a)]. The $X^0$ plateau length is strongly modified by the scattering, whereas the $X^-$ plateau length is unaffected, c.f. Fig.~\ref{densityNORMAL}(a) black versus grey curve. The attractive Coulomb interaction between the QD hole and the Fermi reservoir electrons forms an indirect exciton at the QD position, lowering the ground state difference between the initial and final state of the Fermi reservoir by $\sim G_\text{eh}$. Equivalently, the QD electron tunnels into the Fermi reservoir for transition energies $|\varepsilon_\uparrow^f(V_\text{g}) -(\varepsilon_\text{F}-G_\text{eh})| < \Gamma$, where the calculated Fermi energy $\varepsilon_\text{F}=0$ is at $V_0 = 0.448V$. Here, the tunnel-regime is defined for gate voltages $0.48<V_\text{g}<0.495$, which is confirmed by the simulations, cf. Fig. \ref{densityNORMAL}(c). We find furthermore that a finite scattering potential modifies the energy renormalization of the $X^-$ plateau on its right side (of higher $V_g$-values, approaching the $X^{2-}$ regime). For large detunings from the point of electron-hole symmetry (plateau center) towards the $X^{-2}$ charging state, the model breaks down most likely because the p-shell states of the QD are not considered in the present model.

NRG simulations allow us to calculate the QD electron occupation and the electron occupation of the FR at the QD position. Note, that the hole ($\ket{\Downarrow}$) is traced out in the final Hamiltonian.  The electron occupation of the QD ($\langle \hat{n}_\text{QD} \rangle = \sum_\sigma \langle \hat{n}_\sigma \rangle$) shows nicely the charging regimes of the $X^0$ and $X^-$ while the population inversion of the $\ket{\uparrow}$ and $\ket{\downarrow}$ state of the QD ($\langle m_{QD }\rangle= \langle \hat{n}_\uparrow \rangle - \langle \hat{n}_\downarrow \rangle$) displays the decay of the bright exciton into the dark exciton in the final state and is a measure for the dark state population, provided the hole spin state ($\ket{\Downarrow}$) remains preserved. Therefore we have a negative population inversion in the $X^0$, whereas in the $X^-$ both spin states are degenerate and we have an equal occupation in the final state (zero population inversion). The occupation and the population inversion of the $\ket{\uparrow}$ and $\ket{\downarrow}$ state of the FR at the position of the QD is depicted by $\langle n_{\Psi}\rangle=\sum_\sigma \langle n_{\Psi,\sigma}\rangle$ and $\langle m_{\Psi}\rangle=\langle n_{\Psi,\uparrow}\rangle-\langle n_{\Psi,\downarrow}\rangle$, respectively. The FR is in the initial state (no scattering potential) half filled for each spin, resulting in an expectation value of $\langle n_{\Psi}\rangle = 1$. If the scattering potential is switched on, the scattering potential $G_{ee}$ depletes at the $X^-$  charging state the FR at QD-position in the initial state. After absorption, the hole and the electrons in the QD lead to an attractive (repulsive) potential in the $X^0$ ($X^-$) which induces a charge surplus (deficit) of the FR at the position of the QD in contrast to the initial state. In Fig.~\ref{densityNORMAL}(c) we highlight the tunneling regime, where a Fermi edge singularity arises. For the line shapes with $V_g=0.481$-$0.483\,$V, the occupation decreased already significantly to a QD occupation of $\langle \hat{n}_\text{QD}\rangle$=0.20-0.5. The population inversion also decreases, as the dark state becomes close to the Fermi edge; it is in this regime that the final state of the absorption process is the correlated many-body state.
\bibliographystyle{apsrev4-1}

%

\end{document}